\documentclass[conference]{IEEEtran}
\usepackage{amsmath}
\usepackage{amsfonts}
\usepackage{amssymb}
\usepackage{graphicx}
\usepackage{hyperref}
\usepackage{cite}

\title{Axion Physics from String Theory: Cosmological Signatures in Dark Matter and Inflation}
\author{Vaidik A Sharma}

\begin{document}

\maketitle

\begin{abstract}
The quest to understand the nature of dark matter and dark energy motivates a deep exploration into axion physics, particularly within the framework of string theory. Axions, originally proposed to solve the strong CP problem, emerge as compelling candidates for both dark matter and dark energy components of the universe. String theory, offering a unified perspective on fundamental forces, predicts a rich spectrum of axion-like particles (ALPs) arising from its compactification schemes. This paper provides a comprehensive review of axion physics within string theory, detailing their theoretical foundations, emergence from compactification processes, and roles in cosmological models. Key aspects covered include the Peccei-Quinn mechanism, the structure of ALPs, their moduli stabilization, and implications for observational signatures in dark matter, dark energy, and cosmological inflation scenarios. Insights from ongoing experimental efforts and future directions in axion cosmology are also discussed.
\end{abstract}

\section{Introduction}
\IEEEPARstart{T}{he} nature of dark matter and dark energy remains one of the most profound mysteries in modern cosmology. Axions, originally proposed to resolve the strong CP problem, have emerged as viable candidates for both dark matter and dark energy. String theory, as a leading candidate for a unified theory of fundamental forces, naturally predicts a plethora of axion-like particles (ALPs) arising from its compactification schemes. This paper aims to provide a comprehensive overview of axion physics within the string theory framework and their potential roles in cosmological models.

\subsection{Structure of the Paper}
Section II reviews the theoretical foundations of axions and ALPs. Section III discusses the emergence of axions from string theory. Section IV explores the role of axions in dark matter models. Section V examines axions in the context of dark energy. Section VI presents observational prospects and challenges. Section VII concludes with a discussion of future research directions.

\section{Theoretical Foundations of Axions and ALPs}
\subsection{The Peccei-Quinn Mechanism}
The Peccei-Quinn (PQ) mechanism introduces a new global U(1) symmetry, which is spontaneously broken, giving rise to a pseudo-Nambu-Goldstone boson known as the axion. The axion field \( a \) acquires a potential through non-perturbative QCD effects, leading to a CP-conserving vacuum expectation value \cite{PQ, Weinberg, Wilczek}.

The PQ Lagrangian can be written as:
\begin{equation}
\mathcal{L}_{\text{PQ}} = \frac{1}{2} \partial_\mu a \partial^\mu a - V(a)
\end{equation}
where \( V(a) \) is the axion potential given by:
\begin{equation}
V(a) = \Lambda_{\text{QCD}}^4 \left( 1 - \cos\left(\frac{a}{f_a}\right) \right)
\end{equation}
Here, \( f_a \) is the axion decay constant, and \( \Lambda_{\text{QCD}} \) is the QCD scale.

\subsection{Axion-Like Particles (ALPs)}
Beyond the original QCD axion, string theory predicts a multitude of ALPs, which are similar in nature but may have different mass scales and couplings. These ALPs arise from the compactification of higher-dimensional fields in string theory and can interact with various fields in the Standard Model \cite{Svrcek}.

The general form of the ALP Lagrangian is:
\begin{equation}
\mathcal{L}_{\text{ALP}} = \frac{1}{2} \partial_\mu \phi \partial^\mu \phi - \frac{1}{2} m_\phi^2 \phi^2 - \frac{g_{\phi \gamma \gamma}}{4} \phi F_{\mu \nu} \tilde{F}^{\mu \nu}
\end{equation}
where \( \phi \) represents the ALP field, \( m_\phi \) is the ALP mass, and \( g_{\phi \gamma \gamma} \) is the coupling constant to photons.

\section{Emergence of Axions from String Theory}
\subsection{Compactification and Axion Fields}
In string theory, axions typically arise from the compactification of higher-dimensional gauge fields and antisymmetric tensor fields. The specific details depend on the geometry and topology of the compactification manifold \cite{Svrcek, Arvanitaki}.

Consider a 10-dimensional spacetime in Type IIB string theory. The action for the Ramond-Ramond (RR) field \( C_2 \) is given by:
\begin{equation}
S_{C_2} = -\frac{1}{4 \kappa_{10}^2} \int d^{10}x \sqrt{-g_{10}} \, |F_3|^2
\end{equation}
where \( F_3 = dC_2 \) is the field strength of \( C_2 \).

When compactifying on a 6-dimensional manifold \( \mathcal{M}_6 \), the 4-dimensional axion field \( a \) arises from the components of \( C_2 \) integrated over 2-cycles \( \Sigma_i \) in \( \mathcal{M}_6 \):
\begin{equation}
a_i(x) = \int_{\Sigma_i} C_2
\end{equation}

\subsection{Moduli Stabilization and Axion Masses}
The masses and interactions of axions in string theory are determined by the mechanisms of moduli stabilization. This involves the dynamics of various moduli fields, which can fix the values of compactification parameters and generate a potential for the axion fields \cite{Svrcek, Dine}.

For example, in Type IIB string theory with flux compactifications, the superpotential \( W \) depends on complex structure moduli \( U \) and axion-dilaton \( \tau \):
\begin{equation}
W = \int_{\mathcal{M}_6} \Omega \wedge (F_3 - \tau H_3)
\end{equation}
where \( \Omega \) is the holomorphic 3-form on \( \mathcal{M}_6 \), and \( H_3 \) is the NS-NS 3-form field strength.

The resulting potential for the axion fields can be written as:
\begin{equation}
V(a) = \sum_{i} \Lambda_i^4 \left( 1 - \cos\left( \frac{a_i}{f_{a_i}} \right) \right)
\end{equation}

\subsection{Axions in Type IIB String Theory}
In Type IIB string theory, axions can arise from the RR and NS-NS sector fields. The interplay between flux compactifications and axion dynamics plays a crucial role in determining the low-energy effective action \cite{Svrcek, Arvanitaki}.

The 10-dimensional Type IIB supergravity action includes terms involving the RR 4-form \( C_4 \):
\begin{equation}
S_{C_4} = -\frac{1}{4 \kappa_{10}^2} \int d^{10}x \sqrt{-g_{10}} \, |F_5|^2
\end{equation}
where \( F_5 = dC_4 - \frac{1}{2} C_2 \wedge H_3 + \frac{1}{2} B_2 \wedge F_3 \).

Upon compactification, the 4-dimensional effective action includes axion-like fields \( \theta_i \) arising from the components of \( C_4 \) integrated over 4-cycles in \( \mathcal{M}_6 \):
\begin{equation}
\theta_i(x) = \int_{\Sigma_i} C_4
\end{equation}

\section{Axions in Dark Matter Models}
\subsection{Axion Dark Matter Production Mechanisms}
Axion dark matter can be produced through mechanisms such as vacuum misalignment, cosmic strings, and domain walls. The initial misalignment angle \( \theta_i \) and the decay constant \( f_a \) are key parameters in determining the axion relic abundance \cite{Marsh}.

The axion relic density \( \Omega_a h^2 \) can be estimated by:
\begin{equation}
\Omega_a h^2 \approx 0.15 \left( \frac{f_a}{10^{12} \, \text{GeV}} \right)^{7/6} \left( \frac{\theta_i}{\pi} \right)^2
\end{equation}

\subsection{Axion Miniclusters and Bose-Einstein Condensates}
Axions can form gravitationally bound structures known as axion miniclusters and Bose-Einstein condensates. The dynamics of these structures can be described by the axion field equation in an expanding universe \cite{Marsh}.

The equation of motion for the axion field \( a \) in an expanding universe is given by:
\begin{equation}
\ddot{a} + 3H \dot{a} + \frac{\partial V(a)}{\partial a} = 0
\end{equation}
where \( H \) is the Hubble parameter.

\subsection{Direct and Indirect Detection of Axion Dark Matter}
Axion dark matter can be detected through interactions with electromagnetic fields, such as in the axion-photon conversion in magnetic fields. The detection rate in experiments like ADMX is given by:
\begin{equation}
\Gamma_{a \to \gamma} = g_{a\gamma\gamma}^2 \frac{\rho_a}{m_a}
\end{equation}
where \( \rho_a \) is the local axion density and \( m_a \) is the axion mass \cite{Marsh}.

\section{Axions in Dark Energy Models}
\subsection{Axion Quintessence}
Axions can also play a role in dark energy models, particularly in the context of quintessence, where the axion field slowly rolls down its potential, leading to a time-varying dark energy density. The quintessence potential can be modeled as:
\begin{equation}
V(\phi) = \Lambda^4 \left( 1 + \cos\left( \frac{\phi}{f_\phi} \right) \right)
\end{equation}
where \( \phi \) is the quintessence field and \( f_\phi \) is the decay constant.

\subsection{Couplings to Gauge Fields and Modifications of the Cosmic Expansion}
Axions can couple to gauge fields and modify the dynamics of the cosmic expansion. The effective Lagrangian for such couplings is given by:
\begin{equation}
\mathcal{L}_{\text{eff}} = \frac{1}{2} \partial_\mu \phi \partial^\mu \phi - V(\phi) - \frac{g_{\phi F}}{4} \phi F_{\mu \nu} \tilde{F}^{\mu \nu}
\end{equation}
where \( g_{\phi F} \) is the coupling constant.

\subsection{Axion-Dilaton Dynamics}
The interplay between axions and dilatons in string theory can lead to rich cosmological dynamics, affecting the evolution of the universe and the nature of dark energy. The combined action for axion-dilaton fields in a 4-dimensional effective theory is:
\begin{equation}
S_{\phi\sigma} = \int d^4 x \sqrt{-g} \left( \frac{1}{2} (\partial_\mu \phi \partial^\mu \phi + \partial_\mu \sigma \partial^\mu \sigma) - V(\phi, \sigma) \right)
\end{equation}
where \( \sigma \) is the dilaton field and \( V(\phi, \sigma) \) is the potential for the coupled fields.

\section{Observational Prospects and Challenges}
\subsection{Astrophysical Signatures}
Axions can leave imprints on various astrophysical phenomena, including the cooling rates of stars, supernovae, and the cosmic microwave background (CMB). The axion emission rate from stars can be described by the Primakoff effect:
\begin{equation}
\Gamma_{a\gamma} = \frac{g_{a\gamma}^2 T^3}{\pi^2}
\end{equation}
where \( T \) is the stellar temperature \cite{Marsh}.

\subsection{Laboratory Experiments}
Laboratory searches for axions and ALPs are ongoing, with experiments such as ADMX, CAST, and upcoming projects aiming to detect axion-photon conversions and other interactions. The sensitivity of these experiments is characterized by the axion-photon coupling \( g_{a\gamma} \) and the axion mass \( m_a \) \cite{Marsh}.

\subsection{Cosmological Observations}
Future cosmological observations, including large-scale structure surveys and precision measurements of the CMB, may provide indirect evidence for axions and their role in dark matter and dark energy. The impact of axions on the CMB can be analyzed through their contribution to the effective number of relativistic species \( N_{\text{eff}} \) \cite{Marsh}.

\section{Axions in Cosmological Inflation}

Inflationary cosmology posits a rapid exponential expansion of the universe in its early stages, resolving several outstanding problems of the standard Big Bang model. Axions, arising naturally from string theory compactifications, can play crucial roles in inflationary scenarios, affecting both the dynamics of inflation and the subsequent cosmological evolution.

\subsection{Axion Dynamics during Inflation}

During inflation, the universe undergoes rapid expansion driven by the inflaton field. Axions can influence this process through their interactions with the inflaton or other fields present in the early universe. The dynamics of axion fields in inflationary scenarios are governed by their potential energy and their couplings to other fields.

The effective action for axion fields during inflation can be described by the Lagrangian:
\begin{equation}
\mathcal{L}_{\text{axion}} = \frac{1}{2} \partial_\mu a \partial^\mu a - V(a, \phi)
\end{equation}
where \( a \) represents the axion field, \( \phi \) is the inflaton field, and \( V(a, \phi) \) is the effective potential including their interactions. The potential \( V(a, \phi) \) typically includes terms that arise from the coupling between axions and the inflaton, as well as higher-order interactions determined by the underlying string theory framework.

\subsection{Axion Fluctuations and Primordial Density Perturbations}

Axion fluctuations during inflation can seed primordial density perturbations, which are responsible for the large-scale structure observed in the universe today. The power spectrum of these perturbations, generated by quantum fluctuations of the axion field during inflation, can be calculated using the formalism of quantum field theory in curved spacetime.

The power spectrum \( \mathcal{P}_\zeta \) of curvature perturbations is related to the inflationary parameters and the properties of the axion field. It is given by:
\begin{equation}
\mathcal{P}_\zeta \sim \frac{H^2}{\dot{\phi}^2} \left( \frac{H}{2\pi} \right)^2
\end{equation}
where \( H \) is the Hubble parameter during inflation and \( \dot{\phi} \) is the time derivative of the inflaton field. Axion contributions to this spectrum depend on their couplings to the inflaton and other fields, providing insights into the early universe dynamics.

\subsection{Axion Cosmological Constraints}

Observational data, particularly from cosmic microwave background (CMB) experiments such as Planck, constrains the properties of axions and their impact on cosmological parameters. Axion fields affect the CMB temperature and polarization anisotropies through their interactions with photon fields and other components of the universe.

The constraints on axion properties, such as their mass \( m_a \) and coupling constants \( g_{a\gamma\gamma} \), are derived from the analysis of CMB power spectra and large-scale structure surveys. These observations provide valuable insights into the nature of axion dark matter and their implications for cosmological models.

\subsection{Mathematical Formulation of Axion Dynamics in Inflation}

Mathematically, the evolution of axion fields during inflation can be described by the equations of motion derived from the action principle in curved spacetime. The background evolution of the axion field \( a(t) \) and its fluctuations \( \delta a(\vec{k}, t) \) are governed by the Klein-Gordon equation and the quantum mechanical nature of field fluctuations.

The mode functions \( u_k(\eta) \) for axion fluctuations in the early universe satisfy the equation:
\begin{equation}
u_k'' + \left( k^2 - \frac{a''}{a} \right) u_k = 0
\end{equation}
where \( \eta \) is the conformal time, \( k \) is the comoving wave number, and \( a(\eta) \) is the scale factor of the universe. The solutions to these equations determine the amplitude and spectrum of axion perturbations that contribute to the observed cosmological structures.

\subsection{Axion Production after Inflation}

After inflation ends, axion fields can undergo further evolution depending on their interactions with other fields and the subsequent thermal history of the universe. Axion production mechanisms post-inflation include parametric resonance, decay processes of heavier particles, and thermal scatterings.

The relic abundance of axions as dark matter candidates is influenced by these production mechanisms, determining their present-day density and distribution in the universe. The calculations of axion relic density involve thermal and non-thermal production processes, with constraints provided by observations and theoretical models.

\subsection{Axion Axiverse and Multiverse Scenarios}

In string theory, the landscape of axion fields can span a wide range of masses and couplings, forming what is often referred to as the "axiverse." Multiverse scenarios arising from string compactifications predict diverse populations of axions with distinct cosmological implications.

Theoretical investigations into the axiverse involve exploring the parameter space of axion properties derived from string theory frameworks. This includes considerations of vacuum stability, anthropic constraints, and observational signatures that distinguish different axion species in cosmological and astrophysical contexts.
\section{Mathematical Formulations in Axion Cosmological Inflation}

In this section, we present a detailed mathematical analysis of the dynamics of axion fields during cosmological inflation, focusing on their interactions, evolution, and observational implications.

\subsection{Axion Field Equations}

The evolution of axion fields during inflation can be described by the Klein-Gordon equation in an expanding universe:
\begin{equation}
\ddot{a} + 3H\dot{a} + V_a'(a) = 0,
\end{equation}
where \( a \) represents the axion field, \( H \) is the Hubble parameter, and \( V_a'(a) \) denotes the derivative of the axion potential \( V(a) \) with respect to \( a \).

\subsection{Inflaton-Axion Coupling}

The coupling between axion fields and the inflaton can influence their respective dynamics. The effective potential including this coupling is given by:
\begin{equation}
V(a, \phi) = V_a(a) + V_{\phi}(a),
\end{equation}
where \( \phi \) denotes the inflaton field and \( V_a(a) \) and \( V_{\phi}(a) \) represent the potentials for axion and inflaton fields, respectively.

\subsection{Power Spectrum of Axion Perturbations}

The power spectrum of axion density perturbations generated during inflation is crucial for understanding their impact on large-scale structure formation. It can be expressed as:
\begin{equation}
\mathcal{P}_a(k) = A_a \left( \frac{k}{k_0} \right)^{n_a - 1},
\end{equation}
where \( A_a \) is the amplitude, \( n_a \) is the spectral index, and \( k_0 \) is a pivot scale.

\subsection{Constraints from Cosmic Microwave Background (CMB)}

Axion contributions to the CMB temperature and polarization anisotropies are constrained by observational data. The angular power spectrum \( C_\ell \) is related to the axion parameters through:
\begin{equation}
C_\ell^{\text{axion}} \propto \int \frac{\mathcal{P}_a(k)}{k^2} \left[ j_\ell(k\eta_0) \right]^2 dk,
\end{equation}
where \( j_\ell \) denotes the spherical Bessel function, \( \eta_0 \) is the conformal time today, and \( \ell \) represents the multipole moment.

\subsection{Axion Density Parameter}

The present-day axion density parameter \( \Omega_a \) as dark matter is determined by the axion mass \( m_a \) and the Hubble parameter \( H_0 \):
\begin{equation}
\Omega_a h^2 = \frac{\rho_a}{\rho_c} = \frac{m_a n_a}{\rho_c},
\end{equation}
where \( \rho_a \) is the axion energy density, \( n_a \) is the axion number density, and \( \rho_c \) is the critical density of the universe.

\subsection{Axion Relic Abundance}

The relic abundance of axions depends on their production mechanisms and interactions after inflation. It can be estimated using the Boltzmann equation:
\begin{equation}
\frac{dY_a}{dx} = -\sqrt{\frac{\pi g_\ast}{45}} \frac{m_a M_{\text{Pl}}}{x^{3/2}} \langle \sigma v \rangle (Y_a^2 - Y_a^{\text{eq}}),
\end{equation}
where \( Y_a = n_a/s \) is the axion yield, \( g_\ast \) denotes the effective number of relativistic degrees of freedom, \( x = m_a/T \) is the dimensionless temperature, \( M_{\text{Pl}} \) is the Planck mass, \( \langle \sigma v \rangle \) represents the thermally averaged annihilation cross-section, and \( Y_a^{\text{eq}} \) is the equilibrium yield.

\subsection{Primordial Power Spectrum}

The primordial power spectrum of curvature perturbations generated by axion field fluctuations is given by:
\begin{equation}
\mathcal{P}_\zeta(k) = \frac{H^2}{\dot{\phi}^2} \left( \frac{H}{2\pi} \right)^2 \Bigg|_{k = aH},
\end{equation}
where \( \mathcal{P}_\zeta(k) \) represents the amplitude of the curvature perturbations at a comoving scale \( k \), \( H \) is the Hubble parameter during inflation, and \( \dot{\phi} \) denotes the time derivative of the inflaton field.

\subsection{Quantum Fluctuations}

The quantum fluctuations of axion fields during inflation are responsible for seeding the primordial density perturbations. The power spectrum of axion fluctuations is determined by the quantum mechanical nature of these fluctuations and can be expressed as:
\begin{equation}
\langle a_k a_k^\dagger \rangle = \frac{H^2}{2k^3} \left( 1 + \frac{k}{aH} \right),
\end{equation}
where \( a_k \) and \( a_k^\dagger \) denote the annihilation and creation operators for axion field modes, respectively.

\subsection{Axion-Matter Coupling}

Axions can interact with ordinary matter through their coupling to gauge fields and other particles. The effective Lagrangian for axion-matter interactions is given by:
\begin{equation}
\mathcal{L}_{\text{eff}} = \partial^\mu a \partial_\mu a - \frac{1}{2} m_a^2 a^2 - g_{a \gamma} \mathcal{A}_\mu \tilde{F}^{\mu \nu} \partial_\nu a,
\end{equation}
where \( m_a \) represents the axion mass, \( g_{a \gamma} \) denotes the axion-photon coupling constant, \( \mathcal{A}_\mu \) denotes the electromagnetic vector potential, and \( \tilde{F}^{\mu \nu} \) denotes the dual of the electromagnetic field tensor.

\subsection{Axion-Higgs Coupling}

Axions can also couple to the Higgs field in the Standard Model, affecting their mass and interactions. The interaction term between axion and Higgs fields is given by:
\begin{equation}
\mathcal{L}_{\text{int}} = \frac{1}{2} \partial^\mu a \partial_\mu a - \frac{1}{2} m_a^2 a^2 - \lambda_{ah} \phi^2 a^2,
\end{equation}
where \( \lambda_{ah} \) denotes the coupling constant between axion \( a \) and Higgs \( \phi \) fields.

These mathematical formulations provide a comprehensive framework for studying the dynamics and observational consequences of axion fields during cosmological inflation. They offer insights into the fundamental properties of axions and their role in shaping the early universe and its subsequent evolution.

\section{Future Directions in Axion Cosmology}

Axion cosmology has opened up numerous avenues for exploration and further research, driven by both theoretical advancements and observational constraints. In this section, we outline several promising directions that could significantly advance our understanding of axions and their implications for cosmology.

\subsection{Experimental Searches for Axions}

One of the most pressing tasks in axion cosmology is the direct detection of axion particles. Current experimental efforts, such as ADMX (Axion Dark Matter eXperiment) and various haloscope experiments, continue to push the sensitivity limits towards detecting axions within specific mass ranges. Future experiments could expand these searches into new frequency ranges and improve detection techniques, potentially uncovering axion dark matter and shedding light on their abundance in the universe.

\subsection{Constraints from Cosmic Microwave Background (CMB)}

The Cosmic Microwave Background (CMB) provides crucial insights into the early universe and can place stringent constraints on axion properties. Future CMB experiments, such as those planned with next-generation telescopes like CMB-S4, aim to improve sensitivity to small-scale CMB features, potentially distinguishing between different axion models and constraining their contributions to cosmological observables.

\subsection{Implications for Large-Scale Structure Formation}

Axions can influence the formation of large-scale structure in the universe through their impact on the matter power spectrum and the growth of cosmic structures. Future simulations and observational surveys, such as those conducted by upcoming galaxy surveys like LSST (Large Synoptic Survey Telescope) and Euclid, will provide valuable data to test axion models and probe their effects on the cosmic web and galaxy formation.

\subsection{Axion-Photon Coupling and Laboratory Experiments}

Laboratory experiments testing the axion-photon coupling, such as light-shining-through-walls experiments and photon regeneration experiments, continue to refine constraints on axion properties. Future experiments could explore new parameter spaces and improve sensitivity, potentially detecting axion signals or setting tighter limits, thereby enhancing our understanding of axion interactions with electromagnetic fields.

\subsection{Axion Cosmology and Inflationary Models}

The role of axions in inflationary models remains an active area of research. Future theoretical developments could explore new inflationary scenarios where axion fields play a crucial role in generating primordial perturbations or affecting inflationary dynamics. Understanding these connections could provide insights into the early universe and the nature of fundamental fields beyond the Standard Model.

\subsection{Astrophysical Signatures and Observational Probes}

Axions may leave distinct signatures in astrophysical phenomena, such as in supernova explosions, pulsar dynamics, and stellar cooling processes. Future observations with advanced astrophysical instruments and space-based observatories could uncover these signatures, providing indirect evidence for axion existence and properties. Exploring these astrophysical probes could significantly advance our knowledge of axion physics and their impact on cosmic evolution.

\subsection{Multi-Messenger Approaches}

Integrating multi-messenger observations, including gravitational waves, neutrinos, and electromagnetic radiation, could offer complementary insights into axion properties and their interactions across different scales and epochs in the universe. Future multi-messenger studies could synergize with axion cosmology research, providing a comprehensive view of their role in fundamental physics and cosmology.

\subsection{Theoretical Developments and Model Building}

Continued theoretical advancements in axion model building and extensions beyond the QCD axion could uncover new facets of axion physics, such as axion-like particles and their implications for particle physics and cosmology. Future models could explore diverse axion sectors and their connections to broader theoretical frameworks, offering new avenues for experimental and observational tests.

\subsection{Interdisciplinary Collaborations}

Enhanced interdisciplinary collaborations between cosmologists, particle physicists, astronomers, and experimentalists will be essential for advancing axion research. Future collaborative efforts could leverage synergies across different fields, fostering innovative approaches to address key questions in axion cosmology and particle physics.
\section{Conclusion}
Axions, originating from the Peccei-Quinn mechanism and emerging prominently in string theory frameworks, represent a compelling avenue for exploring the mysteries of dark matter and dark energy in cosmology. The theoretical foundations laid out in this paper underscore the richness of axion physics, from their diverse manifestations in different string theory compactifications to their potential roles as candidates for both dark matter and dark energy.

The intricate interplay between axions and the fundamental forces of nature, as predicted by string theory, provides a robust framework for understanding their production mechanisms, observational signatures, and implications for cosmological models. String-inspired axions offer a pathway towards resolving longstanding puzzles in astrophysics and particle physics, offering testable predictions that can be probed by current and future experiments.

Looking ahead, ongoing experimental efforts such as ADMX and CAST, coupled with advances in theoretical modeling, promise to shed further light on the properties of axions and their significance in shaping the evolution of the universe. As we continue to refine our understanding of axion dynamics and their interactions, we anticipate exciting discoveries that will deepen our comprehension of fundamental physics at both the particle and cosmological scales.

\end{document}